\documentclass[aps,preprint,preprintnumbers,showpacs,nofootinbib]{revtex4}
\usepackage{graphicx}

\begin{document}

\title{Phenomenology of TeV Right-handed Neutrino and the Dark Matter Model}
\author{Kingman Cheung}
\affiliation{
Department of Physics and NCTS, National Tsing Hua University,
Hsinchu, Taiwan, R.O.C.
}
\author{Osamu Seto}
\affiliation{
Institute of Physics, National Chiao Tung University, Hsinchu, Taiwan 300,
R.O.C.}
\date{\today}

\begin{abstract}
In a model of TeV right-handed (RH) neutrino by Krauss, Nasri, and Trodden,
the sub-eV scale neutrino masses are generated via a 3-loop diagram with
the vanishing see-saw mass forbidden by a discrete symmetry, and the TeV mass
RH neutrino is simultaneously a novel candidate for the cold dark matter.
However, we show that with a single RH neutrino it is not possible to
generate two mass-square differences as required by the oscillation data.
We extend the model by introducing one more TeV RH neutrino and
show that it is possible to satisfy the oscillation pattern within
the modified model. After studying in detail the
constraints coming from the dark matter, lepton flavor violation and
the muon anomalous magnetic moment, and the neutrinoless double beta decay,
we explore the parameter space and derive predictions of the model.
Finally, we study the production and decay signatures of the TeV RH neutrinos
at TeV $e^+ e^-/\mu^+ \mu^-$ colliders.  
\end{abstract}
\pacs{}
\preprint{}
\maketitle

\section{Introduction}

One of the most natural way to generate a small neutrino mass is via the
see-saw mechanism \cite{see-saw}.  There are very heavy right-handed neutrinos,
which are gauge singlets of the standard model (SM), 
and so they could have a large
majorana mass $M_R$.  After electroweak symmetry breaking, a Dirac mass term
$M_D$
between the right-handed and the left-handed neutrinos can be developed.  
Therefore, after diagonalizing the neutrino mass matrix, a small majorana
mass $\sim m_D^2/M_R$ for the left-handed neutrino is obtained.  This is
a very natural mechanism, provided that $M_R \sim 10^{11-13}$ GeV.  
One drawback of this scheme is that these right-handed neutrinos are too heavy
to be produced at any terrestrial experiments.  Therefore,
phenomenologically there are not many channels to test the mechanism.
Although it could be possible to get some hints from the neutrino masses
and mixing, it is rather difficult to reconstruct the parameters of the
right-handed neutrinos using the low energy data \cite{kim}.

Another natural way to generate a small neutrino mass is via higher loop
processes, e.g., Zee model \cite{zee}, with some lepton number violating 
couplings.  However, these lepton number violating couplings are also
subject to experimental constraints, 
e.g., $\mu \to e \gamma, \tau \to e\gamma$.
In the Zee model, there are also extra scalars whose masses are of electroweak
scale, and so can be observed at colliders \cite{gl}.

On the other hand, recent cosmological observations have established the
concordance cosmological model where the present energy density consists of
about $73\%$ of cosmological constant (dark energy),
$23\%$ (non-baryonic) cold dark matter, and just $4\%$ of baryons.
To clarify the identity of the dark matter remains a prime open problem
in cosmology and particle physics. Although quite a number of
promising candidates have
been proposed and investigated in detail, other possibilities can never be
neglected.

Recently, Krauss, Narsi, and Trodden \cite{krauss} considered an extension
to the SM, similar to the Zee model, with two additional charged scalar
singlets and a TeV right-handed neutrino.  They showed that with an additional
discrete symmetry the Dirac mass term between the left-handed and right-handed
neutrinos are forbidden and thus avoiding the see-saw mass.  Furthermore, the
neutrino mass can only be generated at three loop level, and sub-eV
neutrino masses can be obtained with the masses of the charged scalars and
the right-handed neutrino of order of TeV.
Phenomenologically, this model is interesting because the TeV right-handed
neutrino can be produced at colliders and could be a dark matter candidate.

In this work, we explore in details the phenomenology of the TeV right-handed
(RH) neutrinos. We shall extend the analysis to three families of left-handed
neutrinos and
explore the region of the parameters that can accommodate the present
oscillation data.  
In the course of our study, we found that the model in Ref. \cite{krauss}
with a single RH neutrino cannot explain the oscillation data, because it
 only gives one mass-square difference.  We extend the model by adding
another TeV RH neutrino, which is slightly heavier than the first one.  We
demonstrate that it is possible to accommodate the oscillation pattern.
We also obtain the relic density of the RH neutrino,
and discuss the possibility of detecting them if they form a 
substantial fraction of the dark matter.  We also study
the lepton number violating processes and the muon anomalous
magnetic moment, and the production at leptonic colliders.
In particular, the pair production of $N_1 N_2, N_2 N_2$
at $e^+ e^-/\mu^+\mu^-$ 
colliders gives rise to very interesting signatures. 
The $N_2$ so produced will decay into $N_1$ plus a pair of charged
leptons inside the detector.  Thus, the signature would be either one or
two pairs of charged leptons plus a large missing energy.

The organization is as follows.  We describe the model in the next section.
In Sec. III, we explore all the phenomenology associated with the
TeV RH neutrino. In Sec. IV, we discuss the signatures in collider experiments.
Section V is devoted to the conclusion.

\section{Review of the Model}

The model considered in Ref. \cite{krauss} has
two extra charged scalar singlets
$S_1, S_2$ and a right-handed neutrino $N_R$.  A discrete $Z_2$
symmetry is imposed on the
particles, such that all SM particles and $S_1$ are even under $Z_2$ but
$S_2, N_R$ are odd under $Z_2$.  Therefore, the Dirac mass term
$\overline{L} \phi N_R$ is forbidden, where $\phi$ is the SM Higgs boson.  The
see-saw mass is avoided.   

In the present work, we extend the model a bit further by adding the second
TeV right-handed neutrino, which also has the odd $Z_2$ parity.
The reason for that is because with only 1 TeV RH neutrino, it is impossible
to obtain two mass-square differences, as required by the oscillation data.
However, with two TeV RH neutrinos it is possible to accommodate
two mass-square differences with the corresponding large mixing angles.
We will explicitly show this result in the next section.
The most general form for the interaction Lagrangian is
\footnote{
In principle, there are terms like $N_1 N_2 \phi$ and
$M N_1 N_2$.  The latter explicitly gives a mixing between the
two RH neutrinos, while the former also gives the mixing after the
Higgs field develops a VEV.  However, the mixing term can be rotated away by
redefining the $N_1$ and $N_2$ fields.  Effectively, the
Lagrangian has the form given in Eq. (\ref{lag}).}
\begin{eqnarray}
{\cal L} &=& f_{\alpha\beta} L^T_\alpha C i \tau_2 L_\beta S^+_1
   + g_{1\alpha} N_1 S_2^+ \ell_{\alpha R}
   + g_{2\alpha} N_2 S_2^+ \ell_{\alpha R} + V(S_1,S_2) + h.c. \nonumber \\
&& + M_{N_1} N^T_1 C N_1 + M_{N_2} N^T_2 C N_2
\label{lag}
\end{eqnarray}
where $\alpha,\beta$ denote the family indices, $C$ is the charge-conjugation
operator, and
$V(S_1,S_2)$ contains a term $\lambda_s (S_1 S_2^*)^2$.  Note that
$f_{\alpha\beta}$ is antisymmetric under interchange of the family indices.
Note that even with the presence of the first term in the Lagrangian
it cannot give rise to the one-loop Zee diagrams for neutrino
mass generation, because there is no mixing term between the Zee charged
scalar $S_1^+$ and the standard model Higgs doublet that can
generate the charged lepton mass.

If the masses of $N_1, N_2, S_1, S_2$ are arranged such that
$M_{N_1} < M_{N_2} < M_{S_1} < M_{S_2}$, $N_1$ would be stable
if the $Z_2$ parity is
maintained.  The $N_1$ could be a dark matter candidate provided that its
interaction is weak enough.  
Also, $N_1,N_2$ must be pair produced or produced associated with
$S_2$ because of the $Z_2$ parity.  
The $N_2$ so produced would decay into
$N_1$ and a pair of charged leptons.  The decay time may be long enough
to produce a displaced vertex in the central detector.
The $S_2$ if produced would also decay into
$N_1, \ N_2$
and a charged lepton.  We will discuss the phenomenology in details
in the next section.

\section{Phenomenology}

\subsection{Neutrino masses and mixings}
The goal here is to find the parameter space of the model in Eq. (\ref{lag})
such that the neutrino mass matrix so obtained can accommodate the
maximal mixing for the atmospheric neutrino, the large mixing angle for the
solar neutrino, and the small mixing angle for $\theta_{13}$ \cite{experiment}:
\begin{eqnarray}
\Delta m_{\rm atm} \approx 2.7 \times 10^{-3} \;{\rm eV}^2 \,, &&
\sin^2 2\theta_{\rm atm} = 1.0 \nonumber \\
\Delta m_{\rm sol} \approx 7.1 \times 10^{-5} \;{\rm eV}^2 \,, &&
\tan^2 \theta_{\rm sol} = 0.45 \nonumber \label{exp-mass-mix} \,, \\
\sin^22\theta_{13} \lesssim 0.1 .
\end{eqnarray}

The three loop Feynman diagram that contributes to the neutrino mass matrix
has been given in Ref. \cite{krauss}.  The neutrino mass matrix
$(M_\nu)_{\alpha\beta}$ is given by
\begin{equation}
\label{mnu}
(M_\nu)_{\alpha\beta} \sim \frac{1}{(4\pi^2)^3} \frac{1}{M_{S_2}}
\lambda_s f_{\alpha \rho} m_{\ell_\rho} g_{\rho} g_\sigma m_{\ell_\sigma}
f_{\sigma \beta}  \label{neutrino mass}
\end{equation}
where $\alpha,\beta$ denote the flavor of the neutrino.
Note that in the Zee model, the neutrino mass matrix entries are proportional
to $f_{\alpha\beta}$ such that only off-diagonal matrix elements are nonzero.
It is well known that the Zee model gives bi-maximal mixings, which have
some difficulties with the large-mixing angle solution of the solar neutrino
\cite{experiment}.
Here in Eq. (\ref{neutrino mass}) we do not have the second Higgs doublet to
give a mixing between the SM Higgs doublet and $S_1^+$, and therefore
the one-loop Zee-type diagrams are not possible.
However, the mass matrix in Eq. (\ref{neutrino mass})
allows for nonzero diagonal
elements, which may allow the departure from the bi-maximal mixings.

The mixing matrix between flavor eigenstates and mass eigenstates is given as
\begin{equation}
U_{\alpha i} = \left(
\begin{array}{ccc}
c_{13}c_{12} & s_{12}c_{13} & s_{13} \\
-s_{12}c_{23}-s_{23}s_{13}c_{12} & c_{23}c_{12}-s_{23}s_{13}s_{12} & s_{23}c_{13} \\
s_{23}s_{12}-s_{13}c_{23}c_{12} & -s_{23}c_{12}-s_{13}s_{12}c_{23} & c_{23}c_{13} \\
\end{array}
\right) ,
\end{equation}
where we have ignored the phases.  The mass eigenvalues are given by
\begin{equation}
U^T M U = M_{\text{diag}} = \text{diag}(m_1,m_2,m_3) .
\end{equation}
The mass-square
differences and mixing angles are related to oscillation data by
\begin{eqnarray}
\Delta m_{sol}^2 &&\equiv \Delta m_{21}^2 = m_2^2-m_1^2 \nonumber \\
\Delta m_{atm}^2 &&\equiv \Delta m_{32}^2 = m_3^2-m_2^2 \nonumber \\
\theta_{sol} &&\equiv \theta_{12} \\
\theta_{atm} &&\equiv \theta_{23} \nonumber \;.
\end{eqnarray}

{}From Eq. (\ref{neutrino mass}) the neutrino mass matrix is rewritten as
\begin{equation}
(M_\nu)_{\alpha\beta} \sim -\frac{\lambda_s}{(4\pi^2)^3M_{S_2}}\left(
\begin{array}{ccc}
(fmg)_e^2 & (fmg)_e(fmg)_{\mu} & (fmg)_e(fmg)_{\tau} \\
(fmg)_e(fmg)_{\mu} & (fmg)_{\mu}^2 & (fmg)_{\mu}(fmg)_{\tau} \\
(fmg)_e(fmg)_{\tau} & (fmg)_{\mu}(fmg)_{\tau} & (fmg)_{\tau}^2 \\
\end{array}
\right) ,\label{mass-matrix}
\end{equation}
where $(fmg)_{\alpha} = \sum_{\rho}f_{\alpha\rho}m_{\ell_{\rho}}g_{\rho}$,
the mass eigenvalues are given by
\begin{eqnarray}
&& m_1 = m_2 = 0, \\
&& m_3 \sim -\frac{\lambda_s}{(4\pi^2)^3M_{S_2}}
\left[(fmg)_e^2+(fmg)_{\mu}^2+(fmg)_{\tau}^2\right] .
\end{eqnarray}
This model obviously cannot explain the neutrino
oscillation data because of the vanishing $\Delta m_{21}^2$.

Hereafter we would like to discuss a possibility to improve this shortcoming.
The reason that this model predicts two vanishing mass eigenvalues is the
proportionality relation in the mass matrix (\ref{mass-matrix}). Therefore
it is necessary to break the proportionality relation.
Although one way to improve the mass matrix might be to add small
perturbations to the original mass matrix, we however
found that this approach cannot resolve the difficulty.
Instead, we consider a modification of the right-handed neutrino sector.
As mentioned before, we employ two TeV RH neutrinos,
the mass matrix (\ref{mass-matrix}) is replaced by
\begin{equation}
(M_\nu)_{\alpha\beta} \sim \frac{1}{(4\pi^2)^3} \frac{1}{M_{S_2}}
\lambda_s \sum_{I=1,2}(f m g_I)_{\alpha}(g_I m f)_{\beta} ,
\label{2-gene neutrino mass}
\end{equation}
where $I$ denotes the two RH neutrinos.

If we assume $(fmg_2)_{\mu} \ll (fmg_1)_e $,
Eq. (\ref{2-gene neutrino mass}) is rewritten as
\begin{equation}
(M_\nu)_{\alpha\beta} \sim -\frac{\lambda_s(fmg_1)_e^2}{(4\pi^2)^3M_{S_2}}
\left(
\begin{array}{ccc}
1+c^2 & w & t+cd \\
w & w^2 & wt \\
t+cd & wt & t^2 +d^2 \\
\end{array}
\right) ,\label{mass-matrix2}
\end{equation}
\begin{eqnarray}
&& w = (fmg_1)_{\mu}/(fmg_1)_e , \nonumber \\
&& t = (fmg_1)_{\tau}/(fmg_1)_e , \nonumber \\
&& c = (fmg_2)_e/(fmg_1)_e ,  \\
&& d = (fmg_2)_{\tau}/(fmg_1)_e ,\nonumber
\end{eqnarray}
and has one zero and two non-zero eigenvalues ;
\begin{equation}
m_{\pm} \sim -\frac{\lambda_s(fmg_1)_e^2}{(4\pi^2)^3M_{S_2}} \lambda_{\pm} ,
\end{equation}
where
\begin{eqnarray}
2\lambda_{\pm} =&& 1+w^2+t^2+c^2+d^2 \nonumber \\
 && \pm\sqrt{(1+w^2+t^2+c^2+d^2)^2-4(d^2+c^2w^2+d^2w^2-2cdt+c^2t^2)},
 \label{lm}
\end{eqnarray}
and each of the mixing angles is given by
\begin{eqnarray}
t_{23} &=& \frac{w(\lambda_+ -c^2-d^2)}{t(\lambda_+-c^2)+cd} , \label{t23} \\
s_{13} &=& \frac{\lambda_+-d^2-tcd}
{\sqrt{(\lambda_+-d^2-tcd)^2+(1+t_{23}^2)w^2(\lambda_+ -c^2-d^2)^2}} ,
\label{s13} \\
c_{12} &=& \frac{1}{c_{13}}\frac{dw}{\sqrt{(c^2+d^2)w^2+(ct-d)^2}} , 
\label{c12}
\end{eqnarray}
where we adopt the normal mass hierarchy. Indeed, we found that
the correct mixing angles could not be realized if we assumed the
inverted mass hierarchy here.
Here $t_{23}\simeq 1$, $s_{13}^2 \ll 1$ imply
$w\simeq t$, $\lambda_+ \gg c^2,d^2$ and $w^2\gg 1$. This means
$t^2\simeq w^2\gg 1, c^2,d^2$. Definitely,
from
\begin{eqnarray}
\sin^22\theta_{13} \simeq \frac{2}{w^2}\left(1- \frac{tcd}{\lambda_+}\right)^2
\left(\frac{2}{1+t_{23}^2}\right) \lesssim 0.1 ,
\end{eqnarray}
we obtain $w^2 \gtrsim 20$. Since Eq. (\ref{c12}) is rewritten as
\begin{equation}
t_{12}^2
\simeq \frac{c^2w^2+(cw-d)^2}{d^2w^2} ,
\end{equation}
where $c_{13}\simeq 1$ and $w\simeq t$ are used, we obtain
\begin{equation}
\frac{c^2}{d^2}\sim \frac{1}{4} ,
\end{equation}
by comparing with Eq.~(\ref{exp-mass-mix}).
{}From the mass-square differences ;
\begin{eqnarray}
\frac{\Delta m_{sol}^2}{\Delta m_{atm}^2} \simeq
\left(\frac{\lambda_-}{\lambda_+}\right)^2 &\simeq&
\left(\frac{2c^2w^2+d^2w^2-2cdw}{4w^4}\right)^2 \nonumber\\
 &\simeq& \left(\frac{3c^2}{2w^2}\right)^2 \sim10^{-2},
\end{eqnarray}
we find  
\begin{equation}
c^2 \sim \pm\frac{4}{3}\left(\frac{w^2}{20}\right) .\label{c}
\end{equation}
Finally, $\Delta m_{atm}^2\simeq m_3^2=m_+^2$ is rewritten as
\begin{eqnarray}
2.7 \times 10^{-3} \;{\rm eV}^2 \simeq
\left(-\frac{40\lambda_s(fmg_1)_e^2}{(4\pi^2)^3M_{S_2}}\right)^2
\left(\frac{w^2}{20}\right)^2 . \label{m+}
\end{eqnarray}
where we used $\lambda_+\simeq 2w^2$.
In the last subsection of this section, we find some parameter space that
leads to correct mixing angles and mass-square 
differences, after considering also
the constraints from the dark matter relic density and lepton flavor violation.

\subsection{Neutrinoless double beta decay}
A novel feature of the majorana neutrino is the existence of neutrinoless
double beta decay, which essentially requires a nonzero entry $(M_\nu)_{ee}$
of the neutrino mass matrix.
The nonobservation of it has put an upper
bound on the size of $(M_\nu)_{ee} \alt 1  $ eV \cite{0nubb}.

In the model with two RH neutrinos, $(M_\nu)_{ee}$ is estimated to be
\begin{eqnarray}
(M_\nu)_{ee} &\sim& -\frac{\lambda_s}{(4\pi^2)^3M_{S_2}}[
(fmg_1)_e^2+(fmg_2)_e^2] \nonumber \\
&\sim& 3 \times 10^{-3}
\left(\frac{1\pm\frac{3}{4}\left(\frac{20}{w^2}\right)}{2}\right) \,\rm{eV}.
\end{eqnarray}
by using Eqs. (\ref{c}) and (\ref{m+}). Thus, we find that this model is
consistent with the current experimental bound.
Such a small $(M_\nu)_{ee}$ may still be within the reach of the GENIUS
neutrinoless double beta decay experiment \cite{genius}.

\subsection{Dark matter: density and detection}

The lightest RH neutrino is stable because of the assumed discrete symmetry.
Here we consider the relic density of the lightest RH neutrino, and the relic
density must be less than the critical density of the Universe.
First of all, we verify that the second lightest RH neutrino is of no
relevance here because of the short decay time.
The heavier RH neutrino
will decay into the lighter one and two right-handed charged leptons,
$N_2 \rightarrow N_1\ell^-_\alpha \ell^+_\beta$ ($\alpha,\beta$ denote
flavors), and its decay width is given by
\begin{eqnarray}
\Gamma_{N_2} &=& \frac{ M_{N_2}}{ 512 \pi^3} |g_{1\beta} g_{2\alpha}|^2
\times \frac{1}{2 \mu_s^2} \Biggr[ 2 (1-\mu_s) (\mu_1 - \mu_s) (
 \mu_1 + \mu_s + \mu_1 \mu_s - 3 \mu_s^2) \log\left( \frac{\mu_s - \mu_1}
                                                          {\mu_s -1} \right)
 \nonumber\\
&& + (1-\mu_1) \mu_s ( 2\mu_1 - 5\mu_s - 5 \mu_1 \mu_s + 6 \mu_s^2 ) -
  2 \mu_1^2 \log \mu_1 \Biggr]
\label{deca}
\end{eqnarray}
where $\mu_1 = M_{N_1}^2/M_{N_2}^2, \mu_s = M_{S_2}^2/M_{N_2}^2$.
In the worst case when $M_{N_2}$ is very close to $M_{N_1}$, say, they
are both of order 1 TeV but differ by 1 GeV only, and we set $g_i \sim 0.1$.
In this case, the decay width is then of order $10^4-10^5 {\rm s}^{-1}$,
i.e., the decay time is still many orders smaller than the age of the present
Universe.  Therefore, the presence of $N_2$
will not affect the relic density of $N_1$.

The relevant interactions for the annihilation is
$N_1 N_1 \to \ell^+_{\alpha R} \ell^-_{\beta R}$
through charged scalar $S_2^+$ exchange.
The corresponding invariant matrix element is given by
\begin{equation}
|\mathcal{M}|^2 = \frac{|g_{1\alpha}g_{1\beta}|^2}{4}
\left[\frac{(2q_1\cdot p_1)2q_2\cdot p_2}{(t-M_{S_2}^2)^2}
+\frac{(2q_2\cdot p_1)2q_1\cdot p_2}{(u-M_{S_2}^2)^2}
-\frac{2M_{N_1}^22p_1\cdot p_2}{(t-M_{S_2}^2)(u-M_{S_2}^2)}\right] ,
\end{equation}
where $q_i$ and $p_i$ are four-momenta of the incoming $N_1$ particles and
the outgoing leptons, respectively.
Then, we obtain
\begin{eqnarray}
2q_1^02q_2^0 \sigma v &=&
\frac{d^3p_1}{(2\pi)^2 2p_1^0}\frac{d^3p_2}{(2\pi)^2 2p_2^0}(2\pi)^2
|\mathcal{M}|^2\delta^{(4)}(q_1+q_2-p_1-p_2) \\
 &=& \frac{1}{8\pi}
 \frac{|g_{1\alpha}g_{1\beta}|^2}{(M_{S_2}^2+\frac{s}{2}-M_{N_1}^2)^2}
 \left[\frac{m_{l\alpha}^2+m_{l\beta}^2}{2}\left(\frac{s}{2}-M_{N_1}^2\right)
\right. \nonumber \\ && \left.
 +\frac{8}{3}\frac{(M_{S_2}^2-M_{N_1}^2)^2+\frac{s}{2}(M_{S_2}^2-M_{N_1}^2)
 +\frac{s^2}{8}}{(M_{S_2}^2+\frac{s}{2}-M_{N_1}^2)^2}
 \frac{s}{4}\left(\frac{s}{4}-M_{N_1}^2\right)
\right] ,
\end{eqnarray}
where $m_{l\alpha}$ is the lepton mass.
We expanded $|\mathcal{M}|^2$ in powers of the 3-momenta of these particles
and integrated over the scattering angle in the second line.
Following ref. \cite{Gondolo}, the thermal averaged annihilation rate is
estimated to be
\begin{eqnarray}
\langle \sigma v\rangle
&=& \left(\frac{M_{N_1}^2T}{2\pi^2}K_2\left(\frac{M_{N_1}}{T}\right)\right)^{-2}
\frac{T}{4(2\pi)^4}\int_{4M_{N_1}^2}^{\infty} ds \sqrt{s-4M_{N_1}^2}
K_1\left( \sqrt{s} /T\right)(2q_1^02q_2^0\sigma v) \nonumber \\
&\simeq& \sum_f\frac{|g_{1\alpha}g_{1\beta}|^2}{32\pi}
\frac{M_{S_2}^4+M_{N_1}^4}{(M_{S_2}^2+M_{N_1}^2)^4}4M_{N_1}^2
\left(\frac{T}{M_{N_1}}\right) \equiv \sigma_0\left(\frac{T}{M_{N_1}}\right),
\label{sigma0:def}
\end{eqnarray}
where $\sum_f$ denotes the summation over lepton flavors, and we have omitted
the contributions from the S-wave annihilation terms, which are
suppressed by the masses of the final state leptons.
%
%
%
%
The relic mass density is given by
\begin{eqnarray}
\Omega_{N_1} h^2 &=& 1.1 \times 10^9
\left.\frac{2(M_{N_1}/T)}{\sqrt{g_*}M_p\langle\sigma v\rangle}\right|_{T_d}
\text{GeV}^{-1}
\end{eqnarray}
where $T_d$ is the decoupling temperature which is determined as
\
\begin{eqnarray}
\frac{M_{N_1}}{T_d} &\simeq&
\ln\left[\frac{0.152}{\sqrt{g_*(T_d)}}M_p\sigma_0M_{N_1}\right]
-\frac{3}{2}\ln\ln\left[\frac{0.152}{\sqrt{g_*(T_d)}}M_p\sigma_0M_{N_1}\right] ,
\label{decoup.temp}
\end{eqnarray}
and $g_*$ is the total number of relativistic degrees of freedom in the thermal
bath \cite{Kolb}.

By comparing with the recent data from WMAP \cite{Bennett:2003bz}, we find
\
\begin{eqnarray}
\Omega_{DM} h^2 = 0.113 =
2.2 \times 10^{12}\left(\frac{M_{N_1}}{10^3\text{GeV}}\right)
\frac{(M_{N_1}/T_d)^2}{\sqrt{g_*}M_p\sigma_0M_{N_1}} . \label{OmegaDM}
\end{eqnarray}
We can calculate $\sigma_0$ from Eqs.(\ref{decoup.temp}) and (\ref{OmegaDM}),
and we obtain
\begin{eqnarray}
\sigma_0
&\simeq&
1.4 \times 10^{-7}\left(\frac{10^2}{g_*(T_d)}\right)^{1/2}
\left(1+0.07\ln\left[\left(\frac{M_{N_1}}{10^3\text{GeV}}\right)
\left(\frac{10^2}{g_*(T_d)}\right) \right] \right) \text{GeV}^{-2} ,
\label{sigma0}
\end{eqnarray}
if we ignore the second term in Eq.(\ref{decoup.temp}). Indeed, we can confirm
the validity of this assumption within about $10\%$ error by using
Eq.(\ref{sigma0}). Actually, Eq.(\ref{decoup.temp}) is evaluated to be
\
\begin{eqnarray}
\frac{M_{N_1}}{T_d} &\simeq&
\ln(2.5\times 10^{13})-\frac{3}{2}\ln\ln(2.5\times 10^{13}) \nonumber \\
&=& 31-5.1 = 26 .
\end{eqnarray}
Our result of $\langle \sigma v\rangle$ is consistent with a
previous estimation \cite{Baltz:2002we}.
Equations (\ref{sigma0:def}) and (\ref{sigma0}) read
\begin{eqnarray}
\sum_f|g_{1\alpha}g_{1\beta}|^2 \simeq
1
\left(\frac{M_{N_1}}{1.3\times 10^2\text{GeV}}\right)^2
\left(\frac{1+M_{S_2}^2/M_{N_1}^2}{1+2}\right)^4
\left(\frac{1+2^2}{1+M_{S_2}^4/M_{N_1}^4}\right) .
\end{eqnarray}
It is obvious that the RH neutrino must be as light as
$\sim 10^2$ GeV and at least one of $g_{1\alpha}$ should be of order of unity,
such that the relic density is consistent with the dark matter measurement.
\footnote
{Krauss et al. \cite{krauss} claimed that $M_{N_R}\sim 1$ TeV and
$g^2\sim 0.1$ is consistent with the dark matter constraint, but in their
rough estimation a numerical factor of $(T_D/M_N)/8 \sim 200$ is missing from
the equation of $\langle \sigma v \rangle$.}
As the mass difference between $M_{S_2}$ and $N_1$ becomes larger,
the upper bound on $M_{N_1}$ becomes smaller provided that 
we keep $g \lesssim 1$.

The detection of the RH neutrinos as a dark matter candidate depends on 
its annihilation
cross section and its scattering cross section with nucleons.  Conventional
search of dark matter employs an elastic scattering signal of the dark
matter with the nucleons.  We do not expect that the $N_R$ dark matter would
be easily identified by this method, given its very mild interaction.
In addition, because of the majorana nature
the annihilation into a pair charged lepton at the present
velocity ($v_{\rm rel}\sim 0$) is also highly suppressed by the small
lepton mass, even in the case of the tau lepton.  However, one possibility
was pointed out by 
Baltz and Bergstrom \cite{Baltz:2002we} that the annihilation
$N_1 N_1 \to \ell^+ \ell^- \gamma$ would not suffer from helicity suppression.
The rate of this process is approximately $\alpha/\pi$ times the
annihilation rate at the freeze-out.  As will be indicated later, the dominant
mode would be $\mu^+ \mu^-\gamma$.  There is a slight chance to observe the
excess in positron, but however the energy spectrum is softened because
of the cascade from the muon decay.  However, the chance of observing
the photon spectrum is somewhat better \cite{Baltz:2002we}.

\subsection{Lepton flavor changing processes and $g-2$}
There are two sources of lepton flavor violation in Eq. (\ref{lag}).  The
first one is from the interaction $f_{\alpha\beta}
L_\alpha^T C i\tau_2 L_\beta S^+_1$.  This one is similar to the Zee model.
(However, the present model would not give rise to neutrino mass terms
in one loop because of the absence of the $S_1^+-\phi$ mixing.)
The flavor violating amplitude of $\ell_\alpha \to \ell_\rho$ via an
intermediate $\nu_\beta$ would be proportional to
$|f_{\alpha\beta} f_{\beta\rho}|$.
The second source is from the term $g_{I\alpha} N_{I} S_2^+ \ell_{\alpha R}$
in the Lagrangian (\ref{lag}).
The flavor violating amplitude of
 $\ell_\alpha \to \ell_\beta$ via an intermediate
$N_{I}$ would be proportional to $|g_{I\alpha} g_{I\beta}|$.   
We apply these two sources to the radiative decays of
$\ell_\alpha \to \ell_\beta \gamma$ and the muon anomalous magnetic moment.

The new contribution to the muon anomalous magnetic moment can be expressed as
\begin{equation}
\Delta a_\mu = \frac{m_\mu^2}{96 \pi^2} \left(
   \frac{|f_{\mu\tau}|^2 + |f_{\mu e}|^2}{M_{S_1}^2 }
 + \frac{6 |g_{1\mu}|^2}{M_{S_2}^2} F_2( M_{N_1}^2/M_{S_2}^2 )
 + \frac{6 |g_{2\mu}|^2}{M_{S_2}^2} F_2( M_{N_2}^2/M_{S_2}^2 )
\right  ) \;,
\end{equation}
where $F_2(x) = (1- 6x + 3x^2 + 2 x^3 - 6 x^2 \ln x)/6(1-x)^4$. The function
$F_2(x) \to 1/6$ for $x\to 0$, and $F_2(0.25) \approx 0.125$.  We naively
put $F_2(x) = 1/6$ for a simple estimate.  Therefore, we obtain
\begin{eqnarray}
\Delta a_\mu &=& 3 \times 10^{-10} \left[
( |f_{\mu\tau}|^2 + |f_{\mu e}|^2 )
\left( \frac{2\times 10^2\; {\rm GeV}}{M_{S_1}} \right)^2 \right. \nonumber \\
&& \left. + \left(|g_{1\mu}|^2+|g_{2\mu}|^2\right)
 \left( \frac{2\times 10^2\; {\rm GeV}}{M_{S_2}} \right)^2
\right ] \; \alt 10^{-9}
\end{eqnarray}
which implies that $f_{23}, f_{21}, g_{1\mu}, g_{2\mu}$
can be as large as $O(1)$ for $O(200$ GeV)
$S_1^+, S_2^+$ without contributing in a significant level to $\Delta a_\mu$.

Among the radiative decays $\mu \to e \gamma$ is the most constrained
experimentally, $B(\mu \to e \gamma) < 1.2 \times 10^{-11}$ \cite{pdg}.
The contribution of the our model is
\begin{equation}
B(\mu\to e\gamma)= \frac{\alpha v^4}{384 \pi} \left[
 \frac{ |f_{\mu\tau} f_{\tau e}|^2}{M_{S_1}^4}
+ \frac{36 |g_{1e} g_{1\mu}|^2}{M_{S_2}^4}\,F^2_2 (M_{N_1}^2/M_{S_2}^2 )
+ \frac{36 |g_{2e} g_{2\mu}|^2}{M_{S_2}^4}\,F^2_2 (M_{N_2}^2/M_{S_2}^2 )
\right ] \;,
\end{equation}
where $v=246$ GeV.
Again we take $F_2(x) = 1/6$ and $O(200$ GeV) mass for $S_1^+, S_2^+$
for a simple estimate.
\begin{eqnarray}
B(\mu\to e\gamma) &=& 1.4 \times 10^{-5} \left[
 ( |f_{\mu\tau} f_{\tau e}|^2) \left(\frac{2\times 10^2 \;{\rm GeV}}{M_{S_1}} \right)^4
+ |g_{1e} g_{1\mu}|^2 \left(\frac{2\times 10^2 \;{\rm GeV}}{M_{S_2}} \right )^4
\right. \nonumber\\
&& \left.
+ |g_{2e} g_{2\mu}|^2 \left(\frac{2\times 10^2 \;{\rm GeV}}{M_{S_2}} \right )^4
 \right ] \; < 1.2 \times 10^{-11} \;,
\end{eqnarray}
which implies that
\begin{equation}
|f_{e\tau} f_{\tau\mu}| < 1\times 10^{-3}, |g_{1e} g_{1\mu}| <1\times 10^{-3},
\; |g_{2e}g_{2\mu}| \; < 1\times 10^{-3}
\end{equation}
This is in contrast to a work by Dicus {\it et al.} \cite{duane}. In their
model, the couplings $g_i$'s are much larger than $f_{ij}$'s.  

\subsection{An example of consistent model parameters}

Here we summarize the constraints from previous subsections, and
illustrate some allowed parameter space.
The prime constraints come from neutrino oscillations.
The maximal mixing and the mass square difference required in the
atmospheric neutrino, and the small $\theta_{13}$ read
\begin{eqnarray}
f_{\tau\mu}m_{\mu}g_{1\mu} \simeq f_{\mu\tau}m_{\tau}g_{1\tau}
\gg f_{e\mu}m_{\mu}g_{1\mu}+f_{e\tau}m_{\tau}g_{1\tau}  
\sim \sqrt{\frac{1}{\lambda_s}\left(\frac{M_{S_2}}{10^2\,\rm{GeV}}\right)}
 \,\rm{MeV} ,
\label{osc1}
\end{eqnarray}
where the terms $f_{\tau e}m_e g_{1e}$ and $f_{\mu e}m_e g_{1e}$
have been omitted because these terms are suppressed by electron mass.
The large mixing angle and the mass square difference required in the
solar neutrino are given by
\begin{eqnarray}
f_{\tau e}m_e g_{2e}+f_{\tau\mu}m_{\mu}g_{2\mu} \simeq 2
(f_{e\mu}m_{\mu}g_{2\mu}+f_{e\tau}m_{\tau}g_{2\tau})   
 \gg f_{\mu e}m_e g_{2e}+f_{\mu\tau}m_{\tau}g_{2\tau} ,
\label{osc2}
\end{eqnarray}
\begin{equation}
\left(\frac{f_{e\mu}m_{\mu}g_{2\mu}+f_{e\tau}m_{\tau}g_{2\tau}}{f_{\tau\mu}m_{\mu}g_{1\mu}}\right)^2 \simeq \frac{2}{3}\times 10^{-1}
\label{mass-def}
\end{equation}
On the other hand, the dark matter constraint requires
at least one of the $g_{1e},\ g_{1\mu},\ g_{1\tau}$ to be of order of unity.
While the muon anomalous magnetic moment does not impose any strong
constraints, lepton flavor violating processes,
especially $B(\mu\to e\gamma)$ gives the following strong constraints
\begin{equation}
|f_{\mu\tau} f_{\tau e}| \lesssim 1\times 10^{-3} \;,
\end{equation}
\begin{equation}
|g_{1e} g_{1\mu}| , |g_{2e} g_{2\mu}| \lesssim 1 \times10^{-3} \;.
\end{equation}

Now, let's look for an example of consistent parameters. From Eq. (\ref{osc1}),
we obtain $|m_{\mu}g_{1\mu}| \simeq |m_{\tau}g_{1\tau}| $, in other words
$|g_{1\mu}|\gg |g_{1\tau}|$, and
\begin{equation}
f_{\tau\mu}\gg f_{e\mu}+f_{\tau e}.
\label{ftm1}
\end{equation}
Since either $g_{1\mu}$ or $g_{1e}$ must be of order of unity from
the dark matter constraint, we take $g_{1\mu} \simeq 1$.
{}From Eqs. (\ref{osc2}) and (\ref{mass-def}) with $g_{2\tau} \simeq 0$,
we obtain
\begin{equation}
f_{\tau\mu} \simeq 2 f_{e\mu}\,, \;\;
|m_{\mu}g_{2\mu}| \gg |m_e g_{2e}|,
\label{ftm2}
\end{equation}
and
\begin{equation}
g_{2\mu}^2 \simeq 8/3 \times 10^{-1}g_{1\mu}^2 \simeq 0.27(g_{1\mu}/1)^2 .
\end{equation}
Equations (\ref{ftm1}) and (\ref{ftm2}) can be rewritten as
\begin{equation}
1 \gg \frac{1}{2}+\frac{f_{\tau e}}{f_{\tau\mu}} ,
\end{equation}
where we find that a mild cancellation between $f_{e\mu}$ and $f_{\tau e}$
is necessary. For instance, $f_{\tau e}/f_{\tau\mu} = -1/3$.
The strong cancellation corresponds to the small $\theta_{13}$.
However, a cancellation with too high accuracy would require a
$\lambda_s$ which is too big by Eq. (\ref{osc1}).
Therefore, one can say that this model predicts a relatively large mixing
in $\theta_{13}$.
Now we obtain an example set of parameters which makes this model workable
and they are
\begin{eqnarray}
&& |g_{1e}|\lesssim 1\times 10^{-3}, |g_{1\mu}| \simeq 1, \,\,\,\, |g_{1\tau}| \simeq 0.06 ,
\nonumber  \\
&& |g_{2e}|\lesssim 2\times 10^{-3} , |g_{2\mu}| \simeq 0.5 , |g_{2\tau}| < 10^{-2},
\nonumber  \\
&& f_{e\mu} \simeq 1\times 10^{-2}, f_{\tau\mu} \simeq 2\times 10^{-2},
f_{e\tau}\sim -f_{e\mu}. \label{fit}
\end{eqnarray}

\section{Production at $e^+ e^-,\mu^+ \mu^-$ colliders}

The decay of $N_2$ may have an interesting signature,
a displaced vertex, in colliders.  Depending on the parameters, $N_2$ could
be able to travel a typical distance, e.g. mm, in the detector
without depositing any kinetic energy, and suddenly decay into $N_1$
and two charged leptons.  The signature is very striking.  

The $N_1 N_1,\,N_2 N_2$ and $N_1 N_2$ pair
can be directly produced at $e^+ e^-$ colliders.  
The differential cross section for $e^+ e^- \to N_I N_I,\,I=1,2$,
is given by
\begin{equation}
\frac{d\sigma}{d\cos \theta}(e^+ e^- \to N_I N_I) =
\frac{g_{Ie}^4 }{256 \pi}\, \frac{\beta_{I} }{s} \, \left[
  \frac{(t-M_{N_I}^2)^2}{(t- M_{S_2}^2)^2 }
+ \frac{(u-M_{N_I}^2)^2}{(u- M_{S_2}^2)^2 }
- \frac{2 M_{N_I}^2 s}{(t-M_{S_2}^2) (u- M_{S_2}^2) } \right ] \;,
\end{equation}
where $\beta_I = \sqrt{ 1- 4 M_{N_I}^2/s}$, $t= M_{N_I}^2
- s/2(1-\beta_I \cos\theta)$, $u= M_{N_I}^2 - s/2(1+\beta_I
\cos\theta)$.  The total cross section
is obtained by integrating over the angle $\theta$:
\begin{eqnarray}
\sigma(e^+ e^- \to N_I N_I) &=&
\frac{g_{Ie}^4 }{64 \pi s}\, \frac{ 2 (x_I-x_s)^2 + x_s}{ -2 x_I^3
+ x_I^2  ( 6 x_s +1) - 2 x_I x_s ( 3x_s + 2 ) +
x_s( 1+ x_s)( 1 + 2 x_s)} \nonumber \\
&& \hspace{-0.7in} \times \left[
\beta_I ( -2 x_I + 2 x_s +1) + 2( (x_I-x_s)^2 + x_s) \log\left(
\frac{ 2x_I - 2x_s + \beta_I -1}{2 x_I - 2x_s - \beta_I -1 }\
 \right ) \right ]
\nonumber
\end{eqnarray}
where $x_I = M_{N_I}^2/s$ and $x_s=M_{S_2}^2/s$.  
For $N_1 N_2$ production the differential cross section
is given by
\begin{eqnarray}
\frac{d\sigma}{d\cos \theta}(e^+ e^- \to N_1 N_2) &=&
\frac{|g_{1e} g_{2e}|^2 }{128 \pi}\, \frac{\beta_{12} }{s} \, \left[
  \frac{(t-M_{N_1}^2)(t-M_{N_2}^2) }{(t- M_{S_2}^2)^2 }
+ \frac{(u-M_{N_1}^2)(u-M_{N_2}^2)}{(u- M_{S_2}^2)^2 } \right.
\nonumber  \\
&& \left.- \frac{2 M_{N_1} M_{N_2}s} {(t-M_{S_2}^2) (u- M_{S_2}^2)}
\right ] \;,
\end{eqnarray}
and the integrated cross section is
\begin{eqnarray}
\sigma(e^+ e^- \to N_1 N_2) &=&
\frac{|g_{1e} g_{2e}|^2 }{128 \pi}\, \frac{\beta_{12} }{s} \,
\nonumber \\
&& \hspace{-1in} \times\frac{4}{\beta_{12} s (-1 +x_{1} +x_{2} - 2 x_s)
                        (-1 +\beta_{12} +x_{1} +x_{2} - 2 x_s)
                        ( 1 +\beta_{12} -x_{1} -x_{2} + 2 x_s)} \nonumber\\
&&\hspace{-1in} \times \Biggr\{
\beta_{12} s (-1 + x_{1} +x_{2} - 2 x_s) (-1 + \beta_{12}^2 + 2 x_{1}
-x_{1}^2 + 2 x_{2} - 6 x_{1} x_{2} - x_{2}^2  \nonumber \\
&& \hspace{-0.7in}
- 4 x_s + 8 x_{1} x_s
+ 8 x_{2} x_s - 8 x_s^2 ) \nonumber \\
&&\hspace{-0.9in} + s (  2 \sqrt{ x_{1} x_{2}}
  + ( x_{1} + x_{2}) (x_{1}+x_{2}-4x_s -1) + 2x_s(2x_s+1) ) \nonumber\\
&&\hspace{-0.7in}  
\times ( \beta_{12}^2 - (-1 + x_{1} + x_{2} -2 x_s)^2 )\log \left(
\frac{-1 -\beta_{12} +x_{1} +x_{2} - 2 x_s}
     {-1 +\beta_{12} +x_{1} +x_{2} - 2 x_s} \right ) \Biggr \}\;,
\end{eqnarray}
where $\beta_{12}= \sqrt{ ( 1- x_{1} - x_{2})^2 - 4 x_{1} x_{2} }$.
The above cross section formulas are equally valid for $\mu^+ \mu^-$
collisions.  Since the constraints from the last section restrict
$g_{1e}$ and $g_{2e}$ to be hopelessly small, we shall concentrate on
using $g_{1\mu}$ and $g_{2\mu}$.

\begin{figure}[th!]
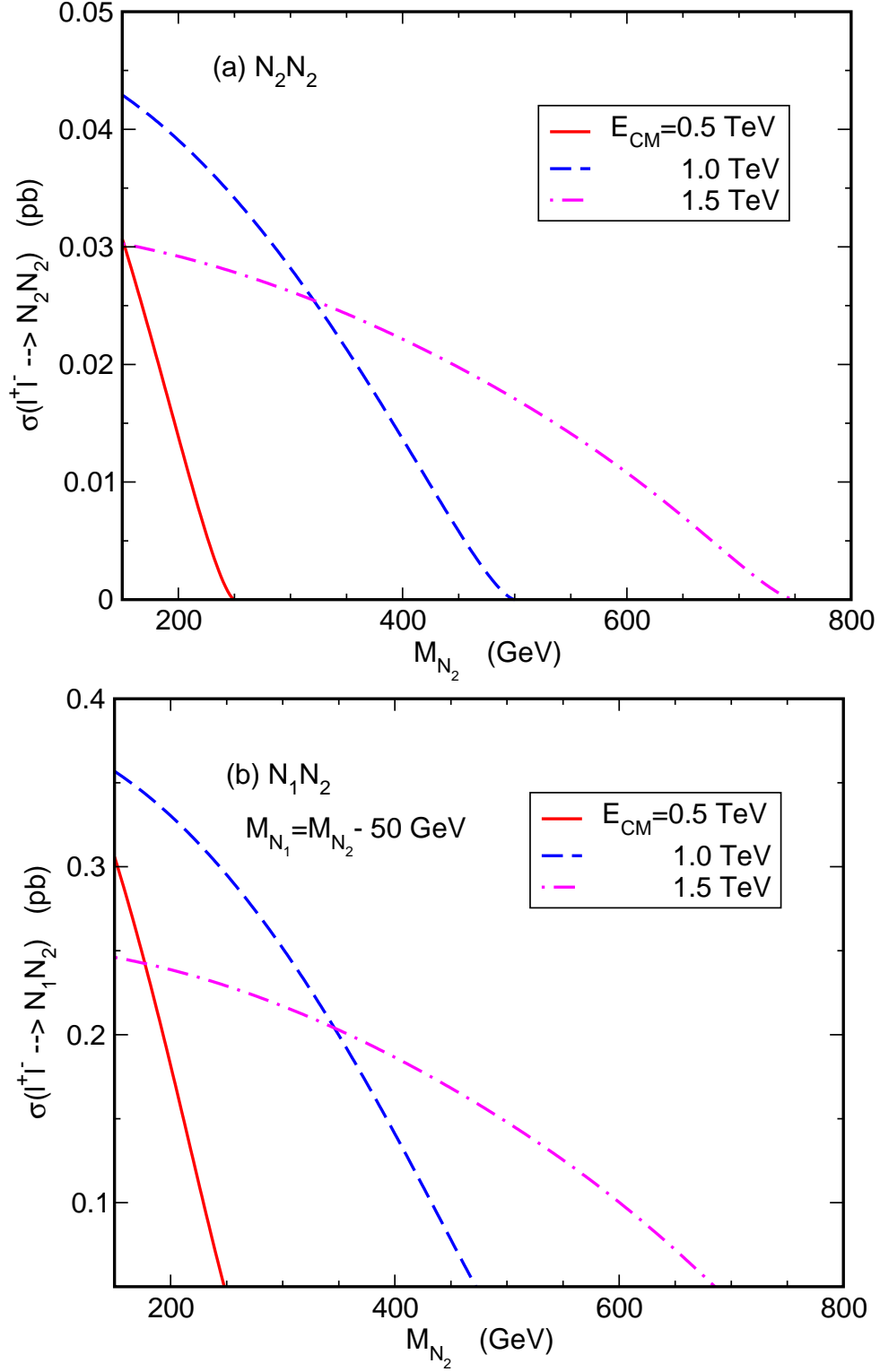

\centering
\includegraphics[height=4in,clip]{fig1a.eps}

\includegraphics[height=4in,clip]{fig1b.eps}
\caption{\small \label{ee-cross}
Production cross sections for (a) $N_2 N_2$ and (b) $N_1 N_2$ pairs for
$\sqrt{s}=0.5,\ 1.0,\ 1.5$ TeV at $l^+ l^-$ collisions.
We have set
$g_{1\mu} = 1,\ g_{2\mu}=0.5$, as suggested by Eq. (\ref{fit}),
$M_{S_2}=500$ GeV, and $M_{N_1} = M_{N_2} - 50 $ GeV.
}
\end{figure}

The production cross sections for
the $N_2 N_2$ and $N_1 N_2$ pairs are given
in Figs. \ref{ee-cross}(a) and (b), respectively,
for $\sqrt{s}=0.5,1,1.5$ TeV and for $M_{N_2}$ from
$150-800$ GeV, and we have set $g_{1\mu}=1,\
g_{2\mu}=0.5$ (see Eq.(\ref{fit})).  In the curve for $N_1 N_2$,
we set $M_{N_1}= M_{N_2} - 50$ GeV.  We are particularly interested in
the $N_1 N_2, N_2 N_2$ production, because of its interesting
signature.

As we have calculated the decay width of $N_2$ in
Eq. (\ref{deca}), the $N_2$ can decay into $N_1$ plus two charged
leptons, either promptly or after traveling a visible distance from the
interaction point.  It depends on the parameters involved, mainly
the largest of $g_{1\beta} g_{2\alpha}$. As seen in Eq. (\ref{fit}) the
largest is $|g_{1\mu} g_{2\mu}| \sim 0.5$, and so the decay of $N_2$ is prompt.
Therefore, in the case of $N_1 N_2$ production, the signature would be
a pair of charged leptons plus missing energies, because the $N_1$'s
would escape the detection.  The charged lepton pair is likely to be on one
side of the event.  In case of $N_2 N_2$ production, the signature
would be two pairs of charged leptons with a large missing energy.
Note that in the case of $N_1 N_1$ production, there are nothing in 
the final state that can be detected.  From Fig. \ref{ee-cross} the production
cross sections are of order $O(10-100)$ fb, which implies plenty of events
with $O(100)$ fb$^{-1}$ luminosity.

One may also consider $S^+_2 S^-_2$ pair production.
The $S_2$ so produced will decay into 
$S^\pm_2 \to N_{1} \ell^\pm_{\alpha R}$ or $N_2 \ell^\pm_{\alpha R}$, where
$\ell_{\alpha} = e, \mu, \tau$.  
However, the
constraints on the parameter space require the mass of $M_{S_2}$ substantially
heavier than $N_1$ and $N_2$, and therefore the $S^+_2 S^-_2$ pair production
cross section is relatively much smaller.

\section{Conclusions}

In this paper, we have discussed a model that explains the small 
neutrino mass and dark matter in the universe at the same time.
Such a model was proposed by Krauss {\it et al.} 
as a modification of Zee model.
However, our study revealed that their original model is unfortunately
not capable of explaining the neutrino oscillation pattern.

We have extended the model by introducing 
another right-handed neutrino.  We succeed in showing
that such an extension is possible to achieve the correct neutrino mixing
pattern.  A prediction of this model is the normal mass hierarchy. 
In addition, the undiscovered mixing angle $\theta_{13}$
is relatively large, 
because of the requirement of a mild cancellation between the parameters for
a small $\theta_{13}$ and a sensible coupling of the 
charged scalar, $\lambda_s$.

The relic density of the lightest right-handed neutrino has also been
revisited. 
Under the constraint by WMAP we found that the mass of the 
right-handed neutrino cannot be as large as TeV but only of order 
$1\times 10^2$ GeV,
after a careful treatment of the calculation.  In addition, other constrains
including the muon anomalous magnetic moment, radiative decay of muon,
neutrinoless double beta decay have also been studied.  With all the 
constraints we are still able to find a sensible region of parameter space.

Finally, our improved model has an interesting signature at leptonic
colliders via pair production of right-handed neutrinos, in particular
$N_1 N_2$ and $N_2 N_2$.  The $N_2$ so produced will decay into $N_1$ plus
two charged leptons.  Thus, the signature is either one or two pairs
of charged leptons with a large missing energy.
Hence, this model can be tested not only by neutrino experiments but also
by collider experiments.

\section*{Acknowledgments}

This research of K.C was supported in part by
the National Science Council of Taiwan R.O.C. under grant no.
NSC 92-2112-M-007-053-.
O.S is supported by the National Science Council of Taiwan under the grant No.
NSC 92-2811-M-009-018-.


\end{document}